# The occurrence of internal gravity waves and volumetric acoustic oscillations in the atmosphere.

## Kochin Alexander V.

Highlights

- The causes of internal gravity waves require additional justification.
- Oscillatory movements with an amplitude of tens of meters are present in the atmosphere in the absence of powerful clouds and fronts.
- Passing the solar terminator causes antiphase pressure and illumination fluctuations.
- The possible process of generating volumetric acoustic oscillations due to the formation of an acoustic resonator in the atmosphere is considered as an alternative to IGW.

Correspondence: Alexander Kochin (amarl@mail.ru)

**Abstract**

Mesoscale wave processes are a transport mechanism for the energy exchange between the troposphere and stratosphere, since the tropopause blocks such an exchange. It is believed that internal gravity waves (IGW) are the main wave process in the atmosphere. However, the explanation of the process of IGW occurrence cannot be considered sufficiently substantiated, because the reasons for the appearance of the selected air parcel, which is affected by the initial disturbance, are not yet clear. The data of the performed experiments on the detection of wave processes shows the presence of noticeable oscillatory movements not only during the passage of fronts or other disturbances, but also in good weather. As an alternative to IGW, I consider a possible process of generating volumetric acoustic oscillations due to the formation of an acoustic resonator in the atmosphere. The probability of formation of such oscillations is high, because they do not require the presence of sharp changes in the parameters of the atmosphere to appear.

**1. Introduction**

Wave motions are important and integral components of dynamic processes in the atmosphere. They form the links between different layers of the atmosphere. They are one of the types of mesoscale turbulence and determine the transport coefficients values on scales of tens and hundreds of kilometers. The tropopause blocks the exchange between the troposphere and the stratosphere due to a sharp decrease in the turbulent transport coefficient. Therefore, the exchange of mass and energy near the tropopause is largely determined by wave processes. Also, wave processes affect the rate of air displacement in the general circulation of the atmosphere, which affects the redistribution of air mass above the Earth's surface. To adequately take into account the contribution of wave processes to the energy and momentum fluxes, it is necessary to know the spatial and temporal characteristics of the perturbation region. However, due to a little number of wave variations studies in the atmosphere and ionosphere, the physical mechanisms that determine the processes of generation and propagation of waves in the atmosphere and ionosphere remain not fully understood. Currently, internal gravity waves (IGW) are considered to be the main wave process that has a noticeable effect on the atmosphere (Alexander, 2004; Gossard, 1975, Holton 2004). However, the possibility of IGW generation raises some doubts, so there is no reason to assume that IGW is the only possible and main type of oscillations. Therefore, the search for alternative types of wave motions seems to be an actual problem.

## 2. Generation of internal gravity waves

Oscillations of any type can be generated either in a hard mode (Fig. 1a) or in a soft mode (Fig. 1b).

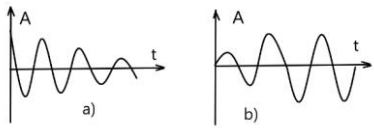

Fig.1. Oscillation generation modes.

In the hard mode, generation starts immediately from the maximum amplitude (Fig. 1a). Then the oscillation amplitude either decreases due to energy dissipation, or remains constant if there is an external energy source that compensates for the losses. Generation in the soft mode starts from zero amplitude and is possible only in the presence of continuous pumping, which provides an increase and maintenance of the oscillation energy due to an external source (Fig. 1b).

The generation of internal gravity waves occurs in a hard mode, as follows from the description of their occurrence (Gossard, 1975; Holton 2004, Tong). It is believed that stable atmospheric stratification contributes to their generation. Therefore, the stratosphere with stable stratification is considered as the main area of IGW formation.

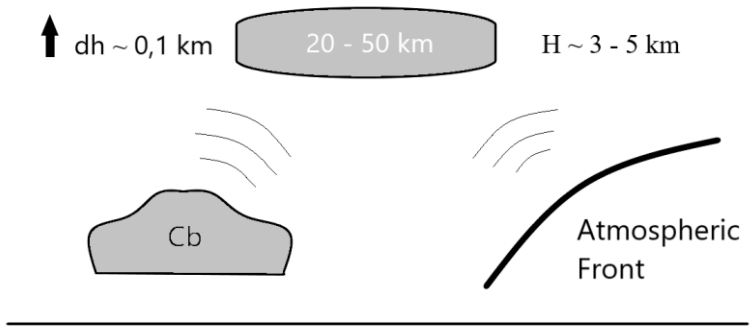

Fig.2. Formation of internal gravity waves. *Cb* - powerful cumulonimbus cloud, *H* - the vertical extent of the volume, *dh* - the initial displacement of the volume.

An external source (a front or a powerful cumulonimbus cloud) creates an impulse that causes an initial displacement($dh$) of some volume as shown in Fig. 2. The size of the air parcel displaced by the initial impulse is not unambiguously determined. Most researchers believe that the dimensions of the air parcel are in the order of a few kilometers in thickness and tens of kilometers long (Alexander, 2004; Gossard, 1975; Kshevetsky, 2015; Kulichkov 2017). We made a paired launch of radiosondes to estimate these sizes (Fig. 3). The results will be discussed in detail below. It turned out that the velocity pulsations remain in-phase on a vertical scale of at least 1.5 km and at a horizontal distance of up to 5 km. The total size of the oscillation region should be several times larger than the region with in-phase velocity oscillations. Thus, the minimum dimensions are 5 km vertically and 20 km horizontally. The mass of such the air parcel in the stratosphere at a height of 15 km with an air density of 0.1 kg/m$^3$ is more than 100 million tons. The amplitude of the initial displacement is tens or hundreds of meters (Fig. 2). The reasons for the sudden displacement of only one specific massive air parcel are difficult to explain. Prior to the occurrence of oscillations, the stratosphere is homogeneous and motionless. The impulse from a powerful cumulonimbus cloud or an atmospheric front affects the entire atmosphere, including the troposphere and stratosphere (Fig. 2). The process that causes the appearance of heterogeneities in the stratosphere with dimensions of tens of kilometers horizontally and a few kilometers vertically has not been described by anyone. If there are no inhomogeneities, then there is no selected object that acquires an additional momentum. Therefore, the existence of oscillations of an isolated air parcel in the stratosphere seems unlikely. There are inhomogeneities in the troposphere, but the oscillations require a vertical component of the added momentum. However, the height difference between irregularities and momentum sources in the troposphere is small, so the vertical component of the added momentum is small. As already noted, the generation of internal gravity waves occurs in a hard mode. Due to damping, the IGW lifetime cannot be long. However, undamped oscillations are recorded for several hours (Borchevkina, 2016), which is also poorly consistent with the mechanism of IGW formation.

The derivation of the the Brunt-Väisälä frequency ratio is done in the hydrostatic approximation, which is valid only for an infinitely small displacement amplitude of an air parcel. It is assumed that the pressure equalization time in the surrounding space is much less than the process time. Equalization of pressures in the medium occurs at the sound speed (Fig. 3).

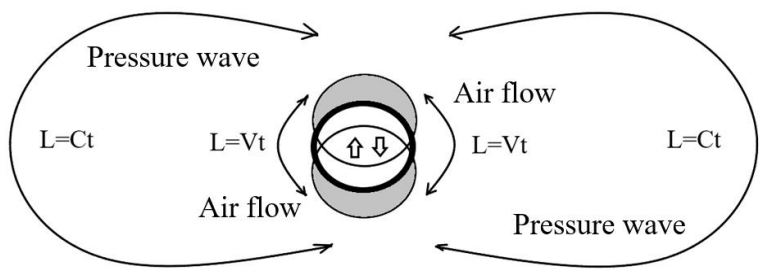

Fig.3. Displacements of a sound wave and an attached air mass during particle oscillation. The added mass of air is highlighted in gray, which is displaced by particle oscillations.

The product of the oscillation period *t* and the speed of sound *C* must be much larger than the size of the oscillation zone *L*.

$$Ct \gg L \qquad (1)$$

For an oscillation region of 20 km, the time to restore hydrostatic equilibrium is about one hour. The period of the Brunt-Väisälä frequency in the stratosphere is 300 seconds. This already limits the scope of applicability of the derivation of the Brunt-Väisälä frequency ratio to a scale of units of kilometers. In addition, when the volume of air is displaced, the vacated volume must be filled by the air from the surrounding space. Then this air from the surrounding space should return to its original place. The shortest path of air displacement corresponds to the perimeter of the oscillation zone.

$$Vt > P \qquad (2)$$

Where *V* is the displacement velocity of the air volume, *t* is the period of oscillation, *P* is the perimeter of the volume of air involved in the oscillatory movement.

During the full period of IGW oscillations (300 seconds in the stratosphere), the path of the added air mass (Fig. 3) cannot be less than the perimeter of the oscillation region. The displacement velocity *V* according to radiosonde measurements is about 0.5 - 1 m/s, which gives a maximum size of 50 - 100 meters. In mountainous areas, velocity up to 5 m/s is observed (Kochin, 2022), so the size can reach 500 meters in isolated cases. In a liquid, the oscillatory velocities are much smaller than in the air, so the expected maximum size of the oscillation region will be even smaller. It should be noted that the process of displacement of the attached air mass leads to damping of vibrations due to the resulting viscous friction. Accordingly, the attenuation of oscillations in the troposphere is stronger than in the stratosphere due to the large value of the turbulent viscosity coefficient. Therefore, displacements of an air particle with a finite amplitude cannot be considered as an adiabatic process.

The above arguments cause doubts about the possibility of generating IGWs; therefore, it is necessary to search for alternative options for the oscillatory processes in the atmosphere.

### 3. Results of observations

To detect oscillatory motions in the atmosphere, an experiment was organized to measure the rate of ascent of two radiosondes launched with 300 seconds delay after each other at 12:00 GMT and at 12:05 GMT on June 5, 2013, accordingly. It is known that the vertical velocity of a radiosonde raised on a shell filled with hydrogen or helium changes insignificantly. There are no reasons for the intrinsic velocity change in oscillatory process, since the amount of gas inside the shell and the elastic properties of the shell remains constant. The lifting force changes smoothly, but cannot change according to the periodic law. Changes in vertical velocity can only be caused by oscillatory movements of the air.

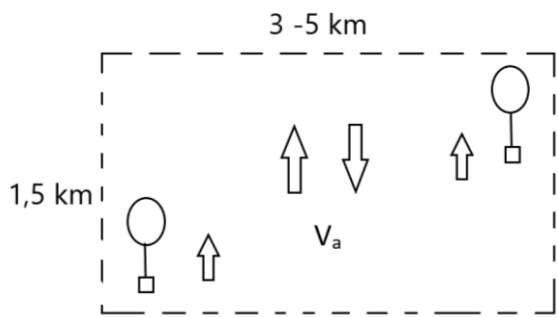

Fig.4. Mutual position of radiosondes launched one after another.

As mentioned above, the radiosondes were released with an interval of 300 seconds (Fig. 4). The ascent rate was 5 m/s, so the vertical distance between the radiosondes was about 1500 meters. The horizontal flow velocity was about 10 m/s, so the horizontal difference varied from 3 to 5 km. The mutual position of the radiosondes is shown in Fig.4.

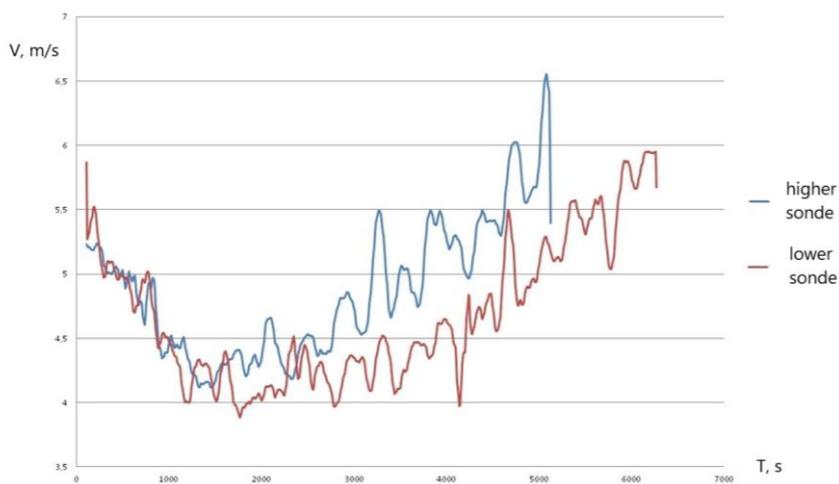

A

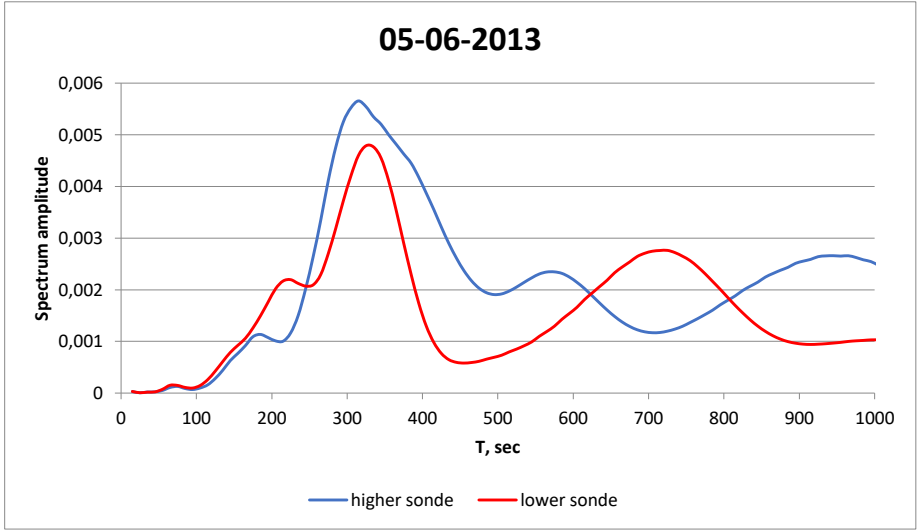

B

Fig.5. A) Vertical velocities of radiosondes launched one after another with 300 seconds delay on 05.06.2013. In the upper figure, X-axis is time in seconds from the time of launch of the second (lower) radiosonde (the data is shown by a red line), Y-axis is vertical ascent rate in m/s. Estimated time of crossing the tropopause, located at an altitude of about 10 km, is 1500 seconds by the first radiosonde (located higher) (blue line) and about 1800 seconds by the second radiosonde (located lower) (red line). B) The spectrum of vertical speed fluctuations of the radiosondes for the entire flight time are shown under the graph with velocities, X-axis is the time in seconds.

Figure 5 shows the records of the vertical rate of ascent of the two radiosondes launched in this experiment and the spectrum of velocity fluctuations. These data demonstrate that the amplitude of velocity fluctuations increases with altitude. Vertical velocity fluctuations occur synchronously. Therefore, there is a process that causes a general oscillation of the entire atmosphere. The period of velocity fluctuations is close to 350 seconds. The vertical oscillatory displacements of the radiosondes from the unperturbed flight path were up to 50 meters.

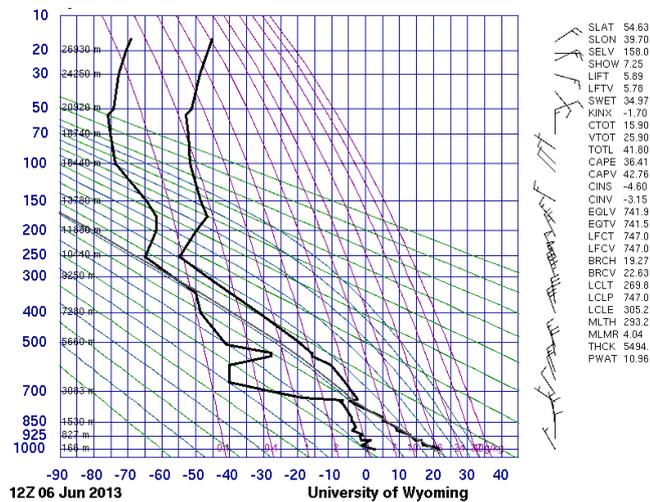

Fig.6. Upper-air sounding data during a paired launch of radiosondes (http://weather.uwyo.edu/upperair/sounding.html). Y-axis is pressure in hPa and height in meters; X axis is temperature (°C). The wind profile is shown by arrows on the right side of the temperature graph.

According to sounding data in the troposphere, a temperature profile was observed with an average vertical gradient of approximately 0.8 °C/100 m (Fig. 6):it is important to note that this gradient was higher than the dry adiabatic gradient (1°K/100 m) at a height of 2 km from the surface of the earth. The temperature profile in the stratosphere corresponded to the isothermal one. There were no clouds. The Brunt-Väisälä frequency period for the measured temperature profile was about 730 seconds in the troposphere and about 300 seconds in the stratosphere. Processing the data of single launches showed that an oscillation period of 350 seconds is often observed, but sometimes a maximum in the oscillation spectrum is observed at periods up to 500 - 800 seconds. Statistical analysis of the duration of the period was not carried out due to the small number of launches.

Observations of atmospheric disturbances passing the solar terminator (ST) are convenient means of verifying various hypotheses. ST is a regular, global source of atmospheric perturbations penetrating the entire atmospheric thickness. ST is characterized by abrupt changes in atmospheric parameters such as temperature and pressure. It is believed that ST causes the formation of internal gravity waves (IGWs). To test the proposed mechanism for the formation of volumetric oscillations, the simultaneous measurement of illumination and atmospheric pressure by a microbarograph at sunrise passing the ST was performed. The measurement data confirmed the presence of synchronous oscillations in the atmosphere. It turned out that antiphase oscillations of atmospheric pressure and illumination are observed with a period of about 350 - 400 seconds (Fig. 7).

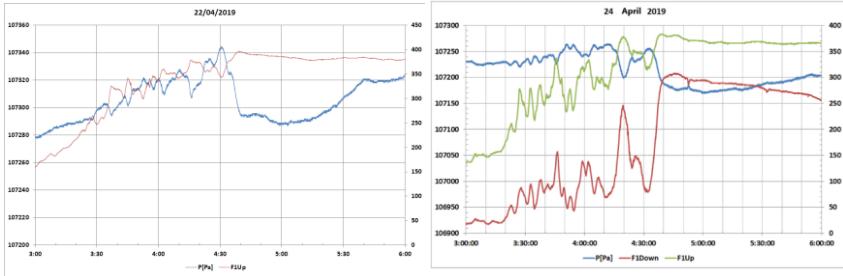

Fig.7. Microbarograph data (blue line) and vertically oriented visible light sensor data (red line). The left Y-axis is the pressure in Pa, the right Y-axis is the light sensor scale in ADC codes, the X-axis is local time in hours. The intensity of the light reflected by the underlying surface is shown by the red line in the right figure.

Illumination fluctuations occur due to changes in the optical mass of the atmosphere, and pressure fluctuations occur due to the acceleration of the center of mass in the process of oscillations. When the center of mass shifts from the equilibrium state to the Earth's surface, the air is compressed and the surface pressure increases. At the same time, the optical mass of the atmosphere decreases, which reduces attenuation (Fig. 8).

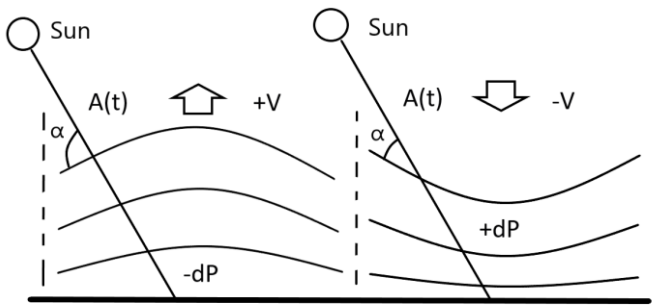

Fig.8. Pressure and illumination changes. $A(t)$ is the displacement amplitude of the layers of the atmosphere during the fluctuation, $dP$ is the fluctuation of surface pressure, $V$ is the vertical velocity of the layers of the atmosphere, $\alpha$ is the angle of sunlight incidence on the atmosphere layers.

Illumination fluctuations and pressure fluctuations should be antiphase, it was confirmed by the experiment. It should be noted that the optical mass of the atmosphere does not change when internal gravity waves occur, since this process causes multidirectional displacement of air in different parts of the atmosphere, which does not change the optical mass of the atmosphere. In addition, light attenuation occurs only in the troposphere, and the stratosphere does not change the intensity of light (Kochin 2021). Therefore, the change in light intensity can only be caused by fluctuations in the troposphere.

The work (Borchevkina, 2016) presents the results of a study of wave processes in the troposphere, which also demonstrate a change of the optical characteristics of the atmosphere when wave processes occur. The measurements were carried out by lidar sounding using a two-wavelength lidar emitting at wavelengths of 532 and 1024 nm (Korshunov, 2007). In these observations, the intensity of the lidar signal from the air layers at a height of 2 and 5 km was measured. It was found that during passing the ST, the intensity fluctuations spectrum of the lidar signal scattered by the atmosphere corresponds to the variations' frequency characteristics of atmospheric parameters (Borchevkina, 2016).

**4. Generation of volumetric acoustic oscillations**

The presence of periodic air movements in the atmosphere in the form of oscillations with different periods has been confirmed by a large number of various measurements (Alexander, 2004; Gossard; 1975; Khaykin, 2015; Kochin, 2014; Kulichkov, 2017). Possible reasons for the occurrence of oscillations are any perturbations of the atmosphere, and not only abrupt changes in the parameters of the atmosphere during passing the terminator, fronts and powerful cumulonimbus clouds. The above-described double launch (Fig. 3) showed the presence of oscillations with an amplitude of up to 50 meters in good weather conditions. Wind flow around orographic inhomogeneities causes a vertical displacement of air masses of comparable magnitude, which can cause oscillations in good weather conditions. Mountain ranges increase the flow disturbances due to the indentation of the relief and cause an increase in the amplitude of oscillations in such areas. With an increment of the wind, disturbances increase. Disturbances in the atmosphere in the form of wind shears, like any inhomogeneity, will also lead to the generation of wave processes.

Vertical displacements of the air mass due to any disturbances lead to a deviation of the vertical pressure profile from the hydrostatic state.

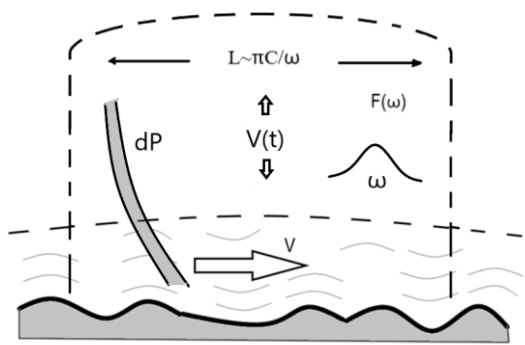

Fig.9. Generation of oscillations due to the deviation of pressure by *dP* (highlighted in gray) from hydrostatic pressure. *V(t)* vertical air flows to restore hydrostatic balance, *V* is the speed of the wind flowing around orographic inhomogeneities, $\omega$ is the frequency of the acoustic resonator, *F($\omega$)* is the response of the acoustic resonator to external influences. *L* is the probable horizontal extent of the oscillation zone; *C* is the speed of sound.

Thus, pressure deviations from hydrostatic pressure (shown in gray in Fig. 9) require subsequent redistribution of mass into atmospheres in order to restore hydrostatic equilibrium. The redistribution of mass leads to the appearance of ordered air flows *V(t)*. However, these air flows do not decrease to zero when the atmosphere reaches hydrostatic equilibrium, these air flows cause oscillatory motions in the vertical direction. The effect is completely analogous to the action of buoyancy forces during the formation of IGWs, including the formation of a certain resonant frequency ω. In this case, the effect of external disturbances is similar to the effect of noise on a resonant system. Like a resonant circuit, the response of an acoustic resonator to white noise is a quasi-harmonic process whose amplitude *U* and phase *φ* vary randomly with time. The frequency of these oscillations is equal to the central frequency of the resonator *ω*. Detailed studies of the occurrence of oscillations in acoustic resonators are already contained in the works of Poincare and Rayleigh (Rayleigh, 1887), but this particular case has not been studied in detail and requires special analysis. The typical size of the oscillating region corresponds to the in-phase movement of air, which corresponds to the distance at which the phase of the oscillation does not change to the opposite due to the propagation time of the wave. For example, at a speed of sound of 330 m/s and a period of about 400 seconds, the size of the in-phase region will reach 50–60 km.

One of the possible options for the formation of an acoustic resonator is shown in Fig. 10.

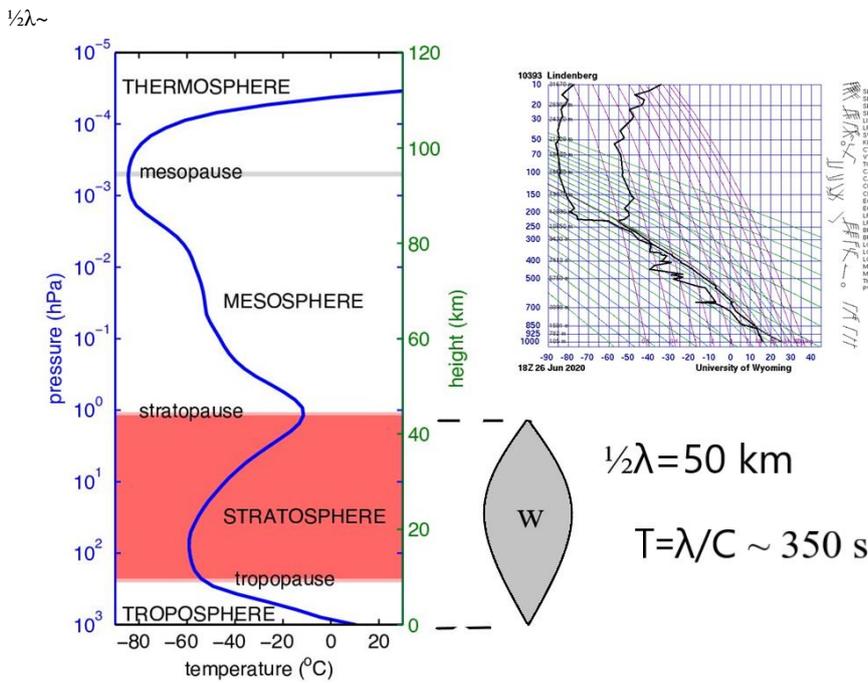

Fig.10. Formation of an acoustic resonator between the Earth's surface and the stratopause and the formation of a standing wave (highlighted in gray). The atmospheric temperature profile (Gerber, 2012) is shown on the left part of the figure. Left Y-axis is pressure (hPa), the right Y-axis is height (km), X-axis is temperature (ºC). An example of upper-air sounding data with wind turn at polar and tropical tropopause heights is shown on the right part of the figure (http://weather.uwyo.edu/upperair/sounding.html). Y- axis is pressure (hPa) and altitude (in meters); X-axis is temperature (ºC). The wind profile is shown by arrows to the right of the temperature graph (http://weather.uwyo.edu/upperair/sounding.html).

Abrupt temperature changes in the atmosphere can be the reason for the formation of an acoustic resonator. The stratopause is located at an altitude of about 50 km and is characterized by a sharp change in temperature (Fig. 10), so it can create a sound-reflecting boundary in the atmosphere. In this case, conditions for the formation of a standing wave between the Earth's surface and the stratopause are created. Half the length of the standing wave (shown in gray in Fig. 10) corresponds to the distance between the boundaries. Accordingly, the wavelength is about 100 km, which corresponds to a wave period of about 350 seconds. In the polar and tropical tropopauses, a 180-degree wind reversal is observed (Cochin, 2022). An example of upper-air sounding data that demonstrates this wind turn is shown in Fig. 10. Displacement of the tropopause height due to a perturbation of the vertical pressure profile can excite a standing wave in the resonator shown in fig.10.

Potentially, oscillations can be amplified due to the thermoacoustic effect. Lord Rayleigh in 1877 in his book "Theory of Sound" formulated the concept of the thermoacoustic effect: "If heat is imparted to the gas at the moment of greatest compression, and heat is taken away at the moment of greatest rarefaction, then this stimulates acoustic vibrations." In the literature, this phenomenon is also called thermal excitation of sound or thermoacoustic oscillations (Basok, 2018; Rott, 1980). When air rises (adiabatic rarefaction), it adiabatically cools, and when it lowers (adiabatic heating), adiabatic heating occurs. A displacement measured in good weather up to 50 meters corresponds to 0.5°K and can potentially amplify vibrations due to the thermoacoustic effect.

Taking all things into consideration, the above arguments about the generation of waves in an atmospheric acoustic resonator are still a hypothesis. However, this hypothesis looks workable and can become the basis for further theoretical and experimental work.

5. **Conclusion**

As the tropopause blocks the exchange of energy between the troposphere and stratosphere, mesoscale wave processes are the main transport mechanism for such exchange. Also, wave processes affect the rate of air displacement in the general circulation of the atmosphere, which affects the redistribution of air mass above the Earth's surface.

It is believed that IGWs are the most paramount wave process in the atmosphere. However, the explanation of the process of IGW occurrence cannot be considered sufficiently substantiated. The reasons for the appearance of a selected volume in the atmosphere, which is affected by the initial perturbation, have not been considered. The derivation of the relationship for the Brunt-Väisälä frequency is made for hydrostatic equilibrium, although the magnitude of the buoyancy force corresponds to the Archimedes equation only in a static state with a hydrostatic pressure distribution. If the object is in motion, then the hydrostatic pressure distribution is violated. Relations (1) and (2) determine the limiting dimensions of objects to which the Archimedes equation is applicable. For example, an elevated inversion in the atmosphere is a layer of warm air hundreds of kilometers long, located above a layer of cold air. Warm air in an elevated inversion remains virtually stationary, although buoyancy forces must act.

The data of the performed experiments on the detection of wave processes shows the presence of noticeable oscillatory movements not only during passing the solar terminator, but also in cloudless weather. The information obtained made it possible to estimate the amplitudes and velocities of air displacement in oscillatory movements, which are tens of meters and up to 1 m/s, accordingly. The period turned out to be close to 350 seconds.

The process of generating volumetric acoustic oscillations due to the formation of an acoustic resonator in the atmosphere can be considered as an alternative to IGW. The probability of formation of such fluctuations will be high, because they do not require the presence of sharp changes in the parameters of the atmosphere to form. To confirm the possibility of implementing the proposed process, it is necessary to conduct additional theoretical and experimental studies.

**Acknowledgments.**

I thank the staff of the Central Aerological Observatory for their help in the work, and the staff of the Dolgoprudnaya (27713) upper-air station for launching radiosondes I express my special gratitude to F. Zagumennov for the development of devices for comparing pressure and illumination, as well as to Dr. M. Khaykin for useful discussions. I also thank my daughter Evgenia Kochina for her help in editing the text.

**References**

1. Alexander, M. J., P. T. May, and J. H. Beres (2004), Gravity waves generated by convection in the Darwin area during the Darwin Area Wave Experiment, *J. Geophys. Res.* doi:10.1029/2004JD004729

2. Basok B., V. V. Gotsulenko. 2018. Self-oscillations excited by the heat sink from the heated gas flow. The journal Proceedings of MIPT 2018. V. 10, № 4.

3. Borchevkina A. Karpov, A.V. A.V. Ilminskaya 2016. The influence of meteorological storms on the parameters of the atmosphere and ionosphere in the Kaliningrad region in 2016. Meteorological Bulletin.

4. Gerber R. & al. (2012). Assessing and Understanding the Impact of Stratospheric Dynamics and Variability on the Earth System. BAMS. DOI: 10.1175/BAMS-D-11-00145.1

5. Gossard E., Hooke W. Waves in the Atmosphere. 1975.

6. Holton J. An Introduction to Dynamic Meteorology. Elsevier. 2004.

7. Khaykin, S. M., A. Hauchecorne, N. Mzé, and P. Keckhut (2015), Seasonal variation of gravity wave activity at midlatitudes from 7 years of COSMIC GPS and Rayleigh lidar temperature observations, Geophys. Res. Lett., 42, doi:10.1002/2014GL062891.

8. Kochin A. 2021. Examination of Optical Processes in The Atmosphere During Upper Air Soundings. Journal of Atmospheric and Oceanic Technology. DOI: 10.1175/JTECH-D-20-0158.1

9. Kochin A. 2022. Correction of Variation due to Non-Hydrostatic Effects The Observed Temperature in Upper-Air Sounding. https://doi.org/10.48550/arXiv.2208.14145

10. Kochin A. 2022. Earth's Rotation Causes Global Atmospheric Circulation. https://doi.org/10.48550/arXiv.2206.14887

11. Kochin A. 2014. Detection of the atmosphere oscillatory motion in the spectra of the electric field and pressure fluctuations. XV International Conference on Atmospheric Electricity, Norman, Oklahoma, USA, 2014

12. Kshevetsky S., Kulichkov N. 2015. Influence of internal gravity waves from convective clouds on atmospheric pressure and spatial distribution of temperature disturbances. *Izvestiya, Atmospheric and Oceanic Physics*. DOI: 10.7868/S000235151501006X

13. Kulichkov S.N., Chunchuzov I.P., Popov O.E., Perepelkin V.G., Golikova E.V., Bush G.A., Repina I.A., Tsybulskaya N.D., Gorchakov G.I. Internal gravity and infrasound waves during a hurricane in Moscow On May 29, 2017 // *Izvestiya, Atmospheric and Oceanic Physics*. doi: 10.31857/S0002-351555232-40

14. Plougonven, R., and F. Zhang (2014), Internal gravity waves from atmospheric jets and fronts, Rev. Geophys., 52, doi:10.1002/2012RG000419.

15. Rayleigh 1887. The theory of sound.

16. Rott, Nikolaus (1980). "Thermoacoustics". *Advances in Applied Mechanics Volume 20*. doi:10.1016/S0065-2156(08)70233-3. ISBN 9780120020201.

17. Tong D. http://www.damtp.cam.ac.uk/user/tong/fluids/fluids3.pdf